# On the weak nematic elasticity

A. I. Leonov[1a] and V.S. Volkov[b]

[a)] *Department of Polymer Engineering, The University of Akron, Akron, OH 44325-0301, USA*
[b)] *Laboratory of Rheology, Institute of Petrochemical Synthesis, Russian Academy of Sciences, Leninsky Pr., 29, Moscow 117912 Russia.*

**Abstract**

The paper considers the general case of incompressible non-classical elasticity with small deformations and rotations. The thermodynamic stability is analysed for free energy density with three rotational degrees of freedom. Although the theory generally predicts the stress to be non-symmetric, the stress tensor can still be considered as symmetrical in the absence of external fields and when the inertia effects of internal rotations and couple stresses are neglected. When the condition of stress tensor symmetry is applied, it results in simplified, "reduced" expressions for the free energy density and stress, which preserve the general stability conditions and allow easy calculations of stress-strain relations and rotations. Using this reduced formulation, a general condition of existence of soft deformations is analysed for the weak nematic elasticity. The reduction procedure is exemplified using the infinitesimal Warner potential derived from a molecular model.

## 1. Introduction

Liquid crystalline (LC) elastomers contain some rigid molecular elements such as mesogenic groups inserted in or side branches attached to the main (backbone) vulcanised polymer chains. As the result, an important mechanical feature of LC elastomers is that these solid materials possess such an additional degree of freedom as internal rotation. Incorporating this effect in a continuum approach allows understanding some unusual mechanisms of deformation for these anisotropic solids.

To describe the macroscopic elasticity of LC elastomers in the equilibrium (static) limit several phenomenological theories have been developed, which included in the elastic free energy the internal rotation terms. De Gennes (1980) first introduced the internal rotations in the free energy density for nematic elastomers, based on symmetry considerations (see also Halperin (1986)). From formal viewpoint, de Gennes (1980) phenomenology is reminiscent the Ericksen (1960) - Leslie (1968) approach for low molecular weight LC nematic liquids. These theories consider only two rotational degrees of freedom, neglecting spin, i.e. the internal rotations about the axis of the one-dimensional isotropy. Brand (1989) analysed effects of electromagnetic fields in rubber nematic elasticity. Terentjev (1993) applied a group-theoretical approach to analyse the linear elasticity and piezoelectric effects in these solids. Brand & Pleiner (1994) and Weilepp & Brand (1996) have also studied the rubber nematodynamics on continuum level (see also Anderson et al (1999), Terentjev & Warner (2001)).

---

[1] Corresponding author.
 Tel.: +1-330-972-5138; fax: +1-330-258-2339
  E-mail address: leonov@uakron.edu (A.I. Leonov).

To derive the free energy for nematic elastomers the elementary molecular theory of rubber elasticity was generalized on the basis of simple models of anisotropic fluctuations in anisotropic environment (e.g. see Abramchuk & Khokhlov (1987), Warner et al (1988), Warner & Wang (1991), Bladon et al (1993), Warner (1999)). When nematic interactions are presented, the chain free energy, even in a Gaussian phantom network approach, has an energetic component, which is a function of the degree of alignment. The first three molecular theories cited above are based on the presentation of the elastic free energy of deformed network in terms of principal stretches, rather than in terms of a set of stretch invariants. Bladon et al (1993) derived the tensor expression for the energy of rubber-like nematic solids. The above molecular models consider only the mono-domain case with uniformly aligned rigid elastomeric groups oriented in uniform external field. Terentjev et al (1996) developed also a molecular theory of nematic rubber elasticity in the presence of non-uniform director field. The cited above molecular models provide a description of some unique phenomena in nematic elastomers, unusual for conventional elasticity. For example, they predict a soft elasticity, first considered by Warner and co-workers (1994), for a class of deformations, which cost no elastic energy when accompanied by director rotation (e.g. see Olmsted (1994) and Warner (1999)).

It should be mentioned that molecular theories, however they important, are mostly based on very specific, particular assumptions, which in many cases might be unwarranted. Thus elaboration of a general continuum approach, which is free of particular assumptions, is desirable.

As the first step in formulation of such a general theory, we present in this paper a continuum theory of weak elasticity for nematic mono-domain solids with three rotational degrees of freedom. Our main objective here is deriving general constitutive equations, which are capable to describe the anisotropic elastic behavior for incompressible nematic solids at small deformations and rotations.

## 2. Kinematics of internal rotations

From macroscopic viewpoint, LC elastomers can be considered as anisotropic solids with internal rotations. Macroscopic theories commonly accept a description of orientation and internal rotations in continuum as evolution of the unit vector $\underline{n}$ called director, and its rigid rotations with angular velocity $\underline{\omega}^I$ (e.g. see Allen & de Silva (1966)). Therefore the speed $\underline{\dot{n}}$ of director is expressed as:

$$\underline{\dot{n}} = \underline{\omega}^I \times \underline{n} = -\underline{\underline{\omega}}^I \cdot \underline{n}, \text{ or } \dot{n}_i = \delta_{ijk}\omega^I_j n_k = -\omega^I_{ik} n_k. \tag{1}$$

Here $\underline{\omega}^I$ and $\underline{\underline{\omega}}^I$ are the vector and related tensor characterizing the speed of internal rotation, $\delta_{ijk}$ is the anti-symmetric unit tensor, and overdot denotes hereafter the time derivative.

For elastic solids of nematic type, an algebraic kinematical relation between the initial value of director $\underline{n}_0$ in non-deformed state and its actual value $\underline{n}$ in the deformed state, equivalently substitutes the rate equation (1). This algebraic relation written via an orthogonal tensor $\underline{\underline{Q}}$ is:

$$\underline{n} = \underline{\underline{Q}} \cdot \underline{n}_0; \quad \underline{\underline{Q}} = \exp(-\underline{\underline{\Omega}}^I). \tag{2}$$

Here $\underline{\underline{\Omega}}^I$ is the anti-symmetric tensor of finite internal rotations. Introducing the vector of internal rotations $\underline{\Omega}^I$ as

$$\Omega^I_{ik} = -\delta_{ijk}\Omega^I_j, \tag{3}$$

represents (2) in the equivalent form:

$$\underline{n} - \underline{n}_0 = [\exp(\underline{\Omega}^I \times)]\underline{n}_0 \equiv \left(\sum_{k=1}^{\infty}\frac{1}{k!}(\underline{\Omega}^I \times)^k\right)\underline{n}_0 \tag{4}$$

Equation (4) shows that $\underline{n} = \underline{n}_0$, when $\underline{\Omega}^I = \underline{\Omega}^I_\parallel \equiv \lambda \underline{n}_0$.

In general case, one can decompose the vector (or corresponding tensor) of internal rotation $\underline{\Omega}^I$ in the sum,

$$\underline{\Omega}^I = \underline{\Omega}^I_\parallel + \underline{\Omega}^I_\perp, \tag{5}$$

where $\underline{\Omega}^I_{\parallel}$ and $\underline{\Omega}^I_{\perp}$ are respectively the vector-components parallel and orthogonal to the initial director $\underline{n}_0$. Nevertheless in spite of (5), the internal spin of finite rotations $\underline{\Omega}^I_{\parallel}$ does not disappear from the equivalent nonlinear kinematical relations (2) or (4).

In the linear case when the internal rotations are small enough, the first relation in (2) takes the approximate form:

$$\underline{n} - \underline{n}_0 \approx -\underline{\underline{\Omega}}^I \cdot \underline{n}_0 = \underline{\Omega}^I \times \underline{n}_0; \quad \left|\underline{\underline{\Omega}}^I\right| = \left|\underline{\Omega}^I\right| \ll 1. \tag{6}$$

In this (and only in this) case, the linear equation (6) can be represented in the equivalent form:

$$\underline{n} - \underline{n}_0 = \underline{\Omega}^I_{\perp} \times \underline{n}_0, \quad \text{or} \quad \underline{\Omega}^I_{\perp} = -\underline{n} \times \underline{n}_0. \tag{7}$$

When the orthogonal tensor $\underline{\underline{Q}}$ is time dependent, differentiating the first relation (2) with respect to time results in: $\underline{\dot{n}} = \underline{\underline{\dot{Q}}} \cdot \underline{n}_0 = \underline{\underline{\dot{Q}}} \cdot \underline{\underline{Q}}^{-1} \cdot \underline{n}$. Comparing this formula with (1) yields:

$$\underline{\underline{\omega}}^I = -\underline{\underline{\dot{Q}}} \cdot \underline{\underline{Q}}^{-1}. \tag{8}$$

In the linear case,

$$\underline{\underline{\omega}}^I \approx \underline{\underline{\dot{\Omega}}}^I, \quad \underline{\omega}^I = \underline{\dot{\Omega}}^I. \tag{9}$$

## 3. Dynamic equations

E. & F. Cosserat (1909) contributed first in the continuum theory of anisotropic solids with internal rotations. According to this theory the orientation of each particle in body can be changed independently of its position. Oseen (1925) developed later a similar theory for liquid crystals.

In these theories, along with the common momentum balance equation

$$\rho \underline{\dot{v}} = \underline{\nabla} \cdot \underline{\underline{\sigma}} + \rho \underline{f}, \tag{10}$$

the rotational degrees of freedom are taken into account by the equation for dynamics of internal angular velocity,

$$\rho \underline{\dot{L}} = \underline{\nabla} \cdot \underline{\underline{\mu}} + \underline{\sigma} + \rho \underline{l}, \quad \sigma_i = \varepsilon_{ijk} \sigma_{jk}. \tag{11}$$

In (10) and (11), $\rho$ is the density, $\underline{v}$ is the velocity, $\underline{\underline{\sigma}}$ is generally non-symmetric stress tensor, $\underline{f}$ is the body force, $\underline{L} = \underline{\underline{I}} \cdot \underline{\omega}^I$ is the internal angular moment, $\underline{\omega}^I$ is the internal angular velocity, $\underline{\underline{I}}$ is the rotational inertia tensor, $\underline{\underline{\mu}}$ is the couple stress tensor, and $\underline{l}$ is the body moment. E. & F. Cosserat (1909) obtained first equation (11) using a variational method. In the Cosserat model with free rotation, internal angular velocity $\underline{\omega}^I$ is considered as an independent variable, consisting of two parts, the velocity of rotation of symmetry axis and the velocity of rotation about the symmetry axis (spin). Thus the stressed state of a continuum with internal rotations is characterized by generally non-symmetric tensors $\underline{\underline{\sigma}}$ and $\underline{\underline{\mu}}$.

The alternative form of the couple-stress equation (11) is:

$$\rho\underline{\underline{\dot{S}}} = 2\underline{\underline{\sigma}}^a + \underline{\underline{m}} + \underline{\underline{l}}, \quad S_{ij} = \delta_{ijk}L_k, \tag{12}$$

where

$$m_{ij} = \delta_{ijk}\nabla_e \mu_{ek}. \tag{13}$$

Here $\underline{\underline{S}}$ is the internal moment of momentum, $\underline{\underline{\sigma}}^a$ is the anti-symmetric part of the stress tensor, and $\underline{l}$ is the body couple. Equation (12) is the balance equation for the internal angular momentum. De Groot and Mazur (1962) who also considered such an equation omitted, however, the angular-momentum flux $F_{ije} = \delta_{ijk}\mu_{ek}$. In the non-polar case, when internal angular momentum, couple stresses, and body moments are negligible, the stress tensor according to (12) is symmetric.

## 4. Incompressible weakly elastic nematics

We derive below constitutive equations for weakly elastic nematic solids, when the elastic strain tensor $\underline{\underline{E}}$, the tensors of internal $\underline{\underline{\Omega}}^I$ and total (body) $\underline{\underline{\Omega}}$ rotations are assumed to be small. Here the tensors $\underline{\underline{E}}$ and $\underline{\underline{\Omega}}$ are defined via the displacement vector $\underline{u}$ by the common formulae of linear elasticity,

$$2\underline{\underline{E}} = \nabla\underline{u} + (\nabla\underline{u})^T; \quad 2\underline{\underline{\Omega}} = \nabla\underline{u} - (\nabla\underline{u})^T. \tag{14}$$

To simplify the analysis we consider below the incompressible case when $tr\underline{\underline{E}} = 0$.

As independent state variables we introduce a scalar parameter $s_r \equiv s - s_0$, characterizing the difference between the values $s$ and $s_0$ of scalar order parameter in deformed and non-deformed states, and two frame invariant tensors, $\underline{\underline{E}}$ and

6$\underline{\underline{\Omega}}^r$ ($\equiv \underline{\underline{\Omega}} - \underline{\underline{\Omega}}^l$). If the state variables $\underline{\underline{E}}$ and $\underline{\underline{\Omega}}^r$ are known, the total internal and body rotations are known separately and the director in the deformed state is found from (6). Additionally, the initial value of director $\underline{n}_0$, which characterizes the uniaxial anisotropy in non-deformed state, is considered in this theory as given.

Following Olmsted (1994) and Warner (1999), we will further neglect the dependence of the free energy on the director gradient, and consider in the following only the "deformational" part of free energy. For nematic elastomers, neglecting the director gradients (i.e. Frank's (1958) elasticity), can be justified for the length scale more than $\sqrt{k/\mu} \approx 10^{-4} cm$. Here $k$ ($\sim 10^{-3} Pa \cdot cm^2$) and $\mu$ ($\sim 10^5 Pa$) are the Frank and rubber elastic moduli, respectively.

Searched within this approach, a general relation for the Helmholtz free energy should be invariant relative to $\underline{n}_0 \rightarrow -\underline{n}_0$ transformation and quadratic in the state variables. Its general form for the weakly elastic case is:

$$\rho F = (G_s/2)s_r^2 + G_{sa}s_r tr(\underline{n}_0\underline{n}_0 \cdot \underline{\underline{E}}) + (G_0/2)tr\underline{\underline{E}}^2 + G_1 tr(\underline{n}_0\underline{n}_0 \cdot \underline{\underline{E}}^2) + G_2 tr^2(\underline{n}_0\underline{n}_0 \cdot \underline{\underline{E}}) \\ - (G_3/2)tr(\underline{\underline{\Omega}}^r)^2 - G_4 tr[\underline{n}_0\underline{n}_0 \cdot (\underline{\underline{\Omega}}^r)^2] - 2G_5 tr(\underline{n}_0\underline{n}_0 \cdot \underline{\underline{E}} \cdot \underline{\underline{\Omega}}^r) \quad (tr\underline{\underline{E}} = 0) \quad (15)$$

Equation (15) is an extended version of equations proposed by De Gennes (1980) and Olmsted (1994) where the Frank contributions are neglected. Those equations written in the vector form, are represented in our notations as:

$$2\rho F = \mu_0 (\underline{n}_0 \cdot \underline{\underline{E}} \cdot \underline{n}_0)^2 + \mu_1 (\underline{n}_0 \times \underline{\underline{E}} \times \underline{n}_0)^2 + \mu_2 (\underline{n}_0 \cdot \underline{\underline{E}} \times \underline{n}_0)^2 + \\ + \alpha_1 (\underline{\underline{\Omega}}^r \times \underline{n}_0)^2 + \alpha_2 \underline{n}_0 \cdot \underline{\underline{E}} \cdot \underline{\underline{\Omega}}^r \times \underline{n}_0 \quad (16)$$

Using the formulae:

$$\underline{n}_0 \times \underline{\underline{E}} \times \underline{n}_0 = \delta_{ejk} n_{0j} E_{ke} n_{0m} \delta_{mes} \quad \text{and} \quad \underline{n}_0 \cdot \underline{\underline{E}} \times \underline{n}_0 = n_{0k} E_{ke} n_{0m} \delta_{mes},$$

equation (16) can be presented in the form (15), where

$$G_0 = \mu_1, \; G_1 = \mu_2/2 - \mu_1, \; G_2 = (\mu_0 + \mu_1 - \mu_2)/2, \; G_4 = \alpha_1/2, \; G_5 = \alpha_2/4, \\ G_s = G_{sa} = G_3 = 0. \quad (17)$$

Far away from the order-disorder phase transition, a nematic elastic solid can be considered as thermodynamically stable, with the stability conditions demanding the quadratic form in (15) to be positive definite. To derive the stability conditions, we introduce a special Cartesian coordinate system, whose one axes, say $x_1$, is directed along the axis of initial director, $\underline{n}_0$. In this coordinate system, where

$$\underline{n}_0 = \{1,0,0\}; \quad \underline{n}_0\underline{n}_0 = \begin{pmatrix} 1 & 0 & 0 \\ 0 & 0 & 0 \\ 0 & 0 & 0 \end{pmatrix}; \quad \underline{\underline{E}} = \hat{\underline{\underline{E}}}; \quad \underline{\underline{\Omega}}^r = \hat{\underline{\underline{\Omega}}}^r,$$

equation (15) for the free energy is represented in the form:

$$\rho \hat{F} = 1/2 G_s s_r^2 + G_{sa} s_r \hat{E}_{11} + (1/2 G_0 + G_1 + G_2)\hat{E}_{11}^2 + 1/2 G_0 (\hat{E}_{22}^2 + \hat{E}_{33}^2 + 2\hat{E}_{23}^2) + G_3(\hat{\Omega}_{23}^r)^2$$
$$+ \sum_{k=2,3} [(G_0 + G_1)\hat{E}_{1k}^2 + 2G_5 \hat{\Omega}_{1k}^r \hat{E}_{1k} + (G_3 + G_4)(\hat{\Omega}_{1k}^r)^2]. \qquad (15a)$$

Since the first five terms and the last two quadratic forms for $k=2$ and $k=3$ are independent, the necessary and sufficient stability conditions are:

$$G_q > 0 ; \; G_0 > 0; \; G_0 + G_1 > 0; \; 1/2 G_0 + G_1 + G_2 - 1/2 G_{sa}^2/G_s > 0; \; G_3 \geq 0;$$
$$(G_0 + G_1)(G_3 + G_4) > G_5^2; \; (G_3 + G_4 > 0). \qquad (18)$$

When using (17), one can easily establish the conditions of stability in terms of the de Gennes coefficients:

$$\mu_0 > 0; \; \mu_1 > 0; \; \mu_2 > 0; \; 4\alpha_1\mu_2 > \alpha_2^2 \; \text{(i.e. } \alpha_1 > 0). \qquad (18a)$$

Note that according to (18) the material parameters $G_{qa}$ and $G_5$ (or $\alpha_2$) are sign indefinite. Some weaker (non-negative) conditions in (18) will be considered later.

The symmetric, $\underline{\underline{\sigma}}^s$, and anti-symmetric, $\underline{\underline{\sigma}}^a$ parts of "extra" stress tensor are calculated using free energy (15) as follows:

$$\underline{\underline{\sigma}}^s = \rho \partial F / \partial \underline{\underline{E}} = G_{sa} s_r \underline{n}_0 \underline{n}_0 + G_0 \underline{\underline{E}} + G_1(\underline{n}_0 \underline{n}_0 \cdot \underline{\underline{E}} + \underline{\underline{E}} \cdot \underline{n}_0 \underline{n}_0) + 2 G_2 \underline{n}_0 \underline{n}_0 tr(\underline{\underline{E}} \cdot \underline{n}_0 \underline{n}_0)$$
$$+ G_5 (\underline{n}_0 \underline{n}_0 \cdot \underline{\underline{\Omega}}^r - \underline{\underline{\Omega}}^r \cdot \underline{n}_0 \underline{n}_0) \qquad (19a)$$

$$\underline{\underline{\sigma}}^a = \rho \partial F / \partial \underline{\underline{\Omega}} = G_3 \underline{\underline{\Omega}}^r + G_4 (\underline{n}_0 \underline{n}_0 \cdot \underline{\underline{\Omega}}^r + \underline{\underline{\Omega}}^r \cdot \underline{n}_0 \underline{n}_0) + G_5 (\underline{n}_0 \underline{n}_0 \cdot \underline{\underline{E}} - \underline{\underline{E}} \cdot \underline{n}_0 \underline{n}_0) . \qquad (19b)$$

Note that the total stress tensor is defined as: $\underline{\underline{\sigma}} = -p\underline{\underline{\delta}} + \underline{\underline{\sigma}}^s + \underline{\underline{\sigma}}^a$. Here $\underline{\underline{\delta}}$ is the unit tensor, and $p$ is an isotropic pressure.

Another equilibrium condition and the corresponding equation for variation of the scalar order parameter $s_r$, are:

$$\rho \partial F / \partial s = 0 ; \; s_r = -(G_{sa}/G_s) tr(\underline{\underline{E}} \cdot \underline{n}_0 \underline{n}_0) . \qquad (19c)$$

It is easy to see that inserting the expression for $s_r$ from (19c) in (15) and (19a), and using instead of $G_2$, another parameter,

$$\hat{G}_2 = G_2 - 1/2 G_{sa}^2 / G_s, \qquad (20)$$

makes it possible to completely exclude the scalar parameter $s_r$ from consideration. Under such a substitution, the free energy and stress will not include the thermodynamic variable $s_r$ and the third stability condition in (18) will be satisfied. Such a substitution is formally possible because both the material parameters, $G_{sa}$ and $G_2$ are sign indefinite. It should be noted, however, that in non-equilibrium situation, the variation of scalar order parameter would produce an additional kinetic equation.

Except the particular case when

$$\hat{G}_2 = -G_1, \qquad (21)$$

the symmetric extra tensor component $\underline{\underline{\sigma}}^s$ is not traceless. The case (21) is, however, of physical significance. As seen from (15a), only in this case the pure nematic contributions in the potential (15), proportional to $G_1$ and $\hat{G}_2$, do not consist the component of elastic strain $\hat{E}_{11}$ directed along the director. It means that this particular case is in a sense related to the "maximal anisotropy". Nevertheless, the condition (21) of the maximal anisotropy can contradict some specific constitutive equations. An example is shown at the end of this paper.

Note that in all the equations of this Section, one can substitute within the same precision the initial value of director $\underline{n}_0$ for the actual one $\underline{n}$. It means that the formulations of free energy in papers [1] and [5] where $\underline{n}$ and $\underline{n}_0$ were respectfully employed, are equivalent. In this substitution, instead of (8a) the reciprocal relation due to (8) might be useful:

$$\underline{n}_0 - \underline{n} \approx \underline{\underline{\Omega}}^I \cdot \underline{n}; \quad \left|\underline{\underline{\Omega}}^I\right| << 1 . \qquad (6a)$$

In such a description, the free energy $F$ in (15) will now depend on the scalar parameter $s$, strain $\underline{\underline{E}}$, relative rotation tensor $\underline{\underline{\Omega}}^r$, and on the actual value of director $\underline{n}$, which is now also variable. Since the director is a unit vector, another Lagrange multiplier $q$ should be introduced when allowing variations of $\underline{n}$. Thus instead of free energy expression $F(s_r, \underline{n}, \underline{\underline{E}}, \underline{\underline{\Omega}}^r)$ shown in (15) where $\underline{n}$ changes to $\underline{n}$, the modified free energy function, $\rho\tilde{F} = \rho F(s_r, \underline{n}, \underline{\underline{E}}, \underline{\underline{\Omega}}^r) - q\underline{n}\cdot\underline{n}$, should be introduced. Then an equilibrium condition, additional to (19), where $\underline{n}$ changes to $\underline{n}$, is of the form: $\rho\partial F/\partial\underline{n} = 2q\underline{n}$. The scalar multiplication this equation by $\underline{n}$ results in:

$q = 1/2 \rho \underline{n} \cdot \partial F / \partial \underline{n}$. This expression shows that the Lagrange multiplier $q$ is a quadratic order of magnitude, i.e. it is an order smaller than the stresses and the variation in order parameter. Although this approach does not provide in equilibrium with a new physical information, it might be fruitful for developing non-equilibrium models, where the algebraic equation (6a) will be changed for an evolution equation for $\underline{n}$.

It should be mentioned that relaxing the assumption of quadratic contribution of elastic strains and internal rotations in the free energy density (15), results in occurring there the additional term, $G_6 tr(\underline{n}_0 \underline{n}_0 \cdot \underline{\underline{E}})$, with a sign indefinite material parameter $G_6$. This term slightly changes the stability conditions, and produces the additional contribution in stress $G_6 \underline{n}_0 \underline{n}_0$, which describes the yield effect. Nevertheless, introducing this term violates the above equivalence between descriptions of the free energy density when using either $\underline{n}$ or $\underline{n}$.

## 5. Reduced free energy and extra stress tensor

In many cases, the non-symmetry of stresses can be neglected. This leads to a highly simplified version of the theory. As seen from (12), it happens when the couple stress tensor $\underline{\underline{\mu}}$ and inertia effects of internal rotations are negligible (here we neglected all the body forces). Unlike suspensions, the inertia effects of internal rotations can be neglected for molecular nematic solids and liquids in a typical mechanical region of frequencies. The couple stresses $\underline{\underline{\mu}}$ are also assumed below to be negligible when external (electrical and magnetic) fields are absent. This is the most important assumption, which within the continuum approach might be verified only when comparing the predictions of the simplified theory of this Section with experiments. It should also be noted that the stress symmetry assumption works very well for the common (non-nematic) elastic solids with small and large deformations.

Using the condition $\underline{\underline{\sigma}}^a = 0$ in (19b) allows us to express the anti-symmetric tensor of relative rotations $\underline{\underline{\Omega}}^r$ via strain tensor $\underline{\underline{E}}$ and initial value of director $\underline{n}_0$ as:

$$\underline{\underline{\Omega}}^r = \Lambda(\underline{\underline{E}} \cdot \underline{n}_0 \underline{n}_0 - \underline{n}_0 \underline{n}_0 \cdot \underline{\underline{E}}); \quad \Lambda = G_5/(G_3 + G_4). \tag{22}$$

Relation (22) has a sense, since due to the stability conditions (18), $G_3 + G_4 > 0$.

Expressing $\underline{\underline{\Omega}}^r$ via the vector relative rotations $\underline{\Omega}^r$ is easy to show that in (22), $\underline{\Omega}^r \cdot \underline{n}_0 = 0$. It means that in the reduced formulation, the spin of internal rotations is equal to the spin of body rotation, i.e. $\underline{\underline{\Omega}}_\parallel^I = \underline{\underline{\Omega}}_\parallel$, and in (22), $\underline{\underline{\Omega}}^r = \underline{\underline{\Omega}}_\perp^r$.

Using (6) one can express the actual value $\underline{n}$ of director as:

$$\underline{n} \approx (\underline{\underline{\delta}} - \underline{\underline{\Omega}}) \cdot \underline{n}_0 + \Lambda[\underline{n}_0 tr(\underline{\underline{E}} \cdot \underline{n}_0 \underline{n}_0) - \underline{\underline{E}} \cdot \underline{n}_0]. \tag{23}$$

Substituting (20) into (15) and (19a) yields the expressions for the reduced free energy $F^r$ and reduced symmetric extra stress $\underline{\underline{\sigma}}^r$ as follows:

$$\rho F^r = \frac{1}{2} G_0^r tr \underline{\underline{E}}^2 + G_1^r tr(\underline{n}_0 \underline{n}_0 \cdot \underline{\underline{E}}^2) + G_2^r tr^2(\underline{n}_0 \underline{n}_0 \cdot \underline{\underline{E}}) \tag{24}$$

$$\underline{\underline{\sigma}}^r = G_0^r \underline{\underline{E}} + G_1^r (\underline{n}_0 \underline{n}_0 \cdot \underline{\underline{E}} + \underline{\underline{E}} \cdot \underline{n}_0 \underline{n}_0) + 2 G_2^r \underline{n}_0 \underline{n}_0 tr(\underline{\underline{E}} \cdot \underline{n}_0 \underline{n}_0), \tag{25}$$

where

$$G_0^r = G_0; \quad G_1^r = G_1 - G_5 \Lambda; \quad G_2^r = \hat{G}_2 + G_5 \Lambda, \tag{26}$$

and $\hat{G}_2$ is determined in (20).

Relations (22)-(24) along with (26) form the closed set of constitutive equations. Remarkable that for the reduced stress there is also the potential relation, $\underline{\underline{\sigma}}^r = \rho \partial F^r / \partial \underline{\underline{E}}$, which holds because of (24) and (25). It is also remarkable that if the strain tensor $\underline{\underline{E}}$ is known, the internal rotations and actual value of director in deformed state are easily established, using Equations (22) and (23), with no additional parameters than those presented in (26). Here the body rotations are established using the compatibility equation.

Note that as in the previous Section, we can substitute within the same precision the initial value of director $\underline{n}_0$ for the actual one $\underline{n}$ in all the equations of this Section except for (23). Because of simplicity, we show below explicitly the related formulae.

Equation (23) will be reduced to:

$$\underline{n}_0 \approx (\underline{\underline{\delta}} + \underline{\underline{\Omega}}) \cdot \underline{n} + \Lambda[\underline{n}\, tr(\underline{\underline{E}} \cdot \underline{nn}) - \underline{\underline{E}} \cdot \underline{n}]. \tag{23a}$$

The modified free energy function $\tilde{F}^r$ is introduced as:

$$\rho \hat{F}^r = \frac{1}{2} G_0^r tr \underline{\underline{E}}^2 + G_1^r tr(\underline{nn} \cdot \underline{\underline{E}}^2) + G_2^r tr^2(\underline{nn} \cdot \underline{\underline{E}}) - q\underline{n} \cdot \underline{n}. \tag{24a}$$

Then the equilibrium condition, additional to the definition of stress $\underline{\underline{\sigma}}^r$ in (25) where $\underline{n}_0$ changed for $\underline{n}$, will be of the form, $\rho \partial \tilde{F}^r / \partial \underline{n} = 2q\underline{n}$. This equation, after scalar multiplication by $\underline{n}$, allows us to determine the value of $q$ as:

$$q = G_1^r tr(\underline{nn} \cdot \underline{\underline{E}}^2) + 2G_2^r tr^2(\underline{nn} \cdot \underline{\underline{E}}).$$

This completely Eulerian approach might be fruitful for developing non-equilibrium models, where the algebraic equation (23a) will be changed for an evolution equation for $\underline{n}$.

## 6. Stability conditions for reduced free energy. Soft and semi-soft deformations

The stability conditions for the reduced free energy (24a) are easily established when using the same coordinate system with relations (16), which represents the free energy in the form:

$$\rho \hat{F}^r = (1/2 G_0^r + G_1^r + G_2^r)\hat{E}_{11}^2 + (G_0^r + G_1^r)(\hat{E}_{12}^2 + \hat{E}_{13}^2) + 1/2 G_0^r (\hat{E}_{22}^2 + \hat{E}_{33}^2 + 2\hat{E}_{23}^2) \quad (27)$$

Since all three terms in (27) are independent, the stability conditions for the reduced free energy are:

$$G_0^r > 0; \quad G_0^r + G_1^r > 0; \quad 1/2 G_0^r + G_1^r + G_2^r > 0. \quad (28)$$

Using relations (26), it is easy to see that the inequalities (28) follow from the general conditions stability (18). In this coordinate system, the components of extra stress tensor are represented as:

$$\begin{aligned}
&\hat{\sigma}_{11}^r = (G_0^r + 2G_1^r + 2G_2^r)\hat{E}_{11}; \quad \hat{\sigma}_{22}^r = G_0^r \hat{E}_{22}; \quad \hat{\sigma}_{33}^r = G_0^r \hat{E}_{33}; \\
&\hat{\sigma}_{12}^r = (G_0^r + G_1^r)\hat{E}_{12}; \quad \hat{\sigma}_{13}^r = (G_0^r + G_1^r)\hat{E}_{13}; \quad \hat{\sigma}_{23}^r = G_0^r \hat{E}_{23}
\end{aligned} \quad (29)$$

It is seen that in the case of maximal anisotropy, when $G_1^r + G_2^r = 0$, the contribution of $\hat{E}_{11}$ in both the reduced free energy and reduced stress will be only due to the isotropic terms, proportional to $G_0^r$. As follows from the second inequality in (28), the sign of $G_1^r$ is still arbitrary in this case too.

Consider now a particular case of elastic nematics, whose material parameters $G_k^r$ from (27) satisfy along with the first and third inequalities in (28), the condition of marginal instability:

$$G_0^r + G_1^r = 0 .\qquad(30)$$

Then Equations (27) and (29) show that applying the shear strains $\hat{E}_{12}$ or $\hat{E}_{13}$ (or both) to a nematic elastic body, initially oriented in the direction $x_1$, causes no occurrence of either corresponding stresses $\hat{\sigma}_{12}^r$ or/and $\hat{\sigma}_{13}^r$, or contribution in the reduced free energy $\hat{F}^r$. These unusual uniform deformations, analyzed in papers by Olmsted (1994), Warner (1999) and others, are called *soft deformations*. This is the special case of soft mode class for anisotropic materials, predicted first by Golubovich and Lubensky (1989). The thermodynamic sense of the soft deformations is getting clear when considering relative rotations described by Equation (22). If both the shears $\hat{E}_{12}$ and $\hat{E}_{13}$ are applied, the tensor of relative rotations is represented due to (22) as:

$$\hat{\underline{\underline{\Omega}}}^r = \Lambda \begin{pmatrix} 0 & -\hat{E}_{12} & -\hat{E}_{13} \\ \hat{E}_{12} & 0 & 0 \\ \hat{E}_{13} & 0 & 0 \end{pmatrix}.\qquad(31)$$

These relative rotations cause the rotation of the director that may be calculated using (23). Thus applying soft deformations causes restructuring the mesogenic groups in nematic solids with spontaneous re-orientation of these groups whose rotational part of free energy is spent for the expense of pure deformation part of free energy. Interestingly, this restructuring theoretically happens with internal re-distribution but no spending of the total free energy. Golubovich and Lubensky (1989) discussed more detailed physics of these soft transitions.

Note that in the nematic "maximal anisotropic" case, when $G_1^r + G^r = 0$, the marginal instability condition (30) in the reduced potential, allows us to express all the coefficients in formulae (24) and (25) via only $G^r$ as:

$$G_1^r = -G_0^r,\ G_2^r = G_0^r\ (G_0^r > 0).$$

Practically, the soft modes represent an idealization of real processes that have *semi-soft* deformations, observed for nematic elastomers (e.g. see Warner (1999) and references there). The semi-soft deformations can be described with a very small positive non-dimensional parameter $\kappa$ as:

$$G_0^r + G_1^r = \kappa G_0^r .\qquad(30a)$$

Because of relation (30a) the shear stresses $\hat{\sigma}_1^r$ and $\hat{\sigma}_1^r$, corresponding to shear strains $\hat{E}_{12}$ and $\hat{E}_{13}$, are very small, as well as their corresponding contribution in the free energy.

## 7. Example: infinitesimal Warner potential

Warner *et al* (1993) using entropy concept derived the following expression of free energy for nematic elastomers:

$$2\rho F^w / G = tr(\underline{\underline{l}}_0 \cdot \underline{\underline{F}} \cdot \underline{\underline{l}}^{-1} \cdot \underline{\underline{F}}^T) \tag{32}$$

Here $\underline{\underline{l}}_0$ and $\underline{\underline{l}}$ are the tensors characterizing anisotropy in initial (non-deformed) and actual (deformed) states, and $\underline{\underline{F}}$ is the strain gradient tensor. In (32):

$$\underline{\underline{l}}(\underline{n}) = l_\perp \underline{\underline{\delta}} + (\Delta l)\underline{nn}; \quad \underline{\underline{l}}^{-1}(\underline{n}) = 1/l_\perp \underline{\underline{\delta}} - \Delta l /(l_\perp l_\parallel)\underline{nn};$$
$$\underline{\underline{l}}_0 = l_\perp^0 \underline{\underline{\delta}} + (\Delta l^0)\underline{n}_0\underline{n}_0; \quad \Delta l^0 = l_\parallel^0 - l_\perp^0; \quad \Delta l = l_\parallel - l_\perp. \tag{33}$$

The parameters of macromolecular chain anisotropy $l_\parallel$ and $l_\perp$ depend on the nematic scalar order parameter *s*. Explicit form of these dependences is different for different models of nematic polymers. Unlike the common assumption that the direction of preferred orientation changes only due to the action of external fields, we assume here that in a mechanical field, parameter *s* can also be slightly changed as described in Section 4. It yields that $l_\parallel = l_\parallel^0 + O(s_r)$ and $l_\perp = l_\perp^0 + O(s_r)$, where $s_r = s - s_0$. Thus taking into account the formulae (15) and (19c), we can extend the Olmsted (1994) calculations of the Warner potential for weakly elastic case to:

$$\frac{\rho F^w}{G} = -\alpha tr^2(\underline{nn} \cdot \underline{\underline{E}}) + tr(\underline{\underline{E}}^2) + \frac{(\Delta l)^2}{2l_\perp l_\parallel}\{tr(\underline{nn} \cdot \underline{\underline{E}}^2) - tr^2(\underline{nn} \cdot \underline{\underline{E}}) - tr[\underline{nn} \cdot (\underline{\underline{\Omega}}^r)^2]\}$$
$$- \frac{\Delta l^2}{l_\perp l_\parallel} tr[\underline{nn} \cdot (\underline{\underline{E}} \cdot \underline{\underline{\Omega}}^r)]; \quad (\Delta l)^2 = (l_\parallel - l_\perp)^2; \quad \Delta l^2 = l_\parallel^2 - l_\perp^2; \quad \alpha = \frac{G_{sa}^2}{G_s G} > 0. \tag{34}$$

To simplify the notations we changed in (34) $\underline{n}_0$ for $\underline{n}$, $l_\parallel^0$ for $l_\parallel$, and $l_\perp^0$ for $l_\perp$.

Comparing (34) with (15), where $s_r$ has been substituted using (19c), one can notice that (34) presents a very particular version of the general potential (15), with the correspondence between material parameters in (15) and those in (34) given as:

$$G_0 = 2G; \quad G_1 = -G_2 = G_4 = G\frac{(l_\parallel - l_\perp)^2}{2l_\parallel l_\perp}; \quad G_3 = 0; \quad G_5 = G\frac{l_\parallel^2 - l_\perp^2}{2l_\parallel l_\perp}. \tag{35}$$

It is easy to see that the material parameters in (35), along with parameter $\alpha$, satisfy the thermodynamic stability conditions (18).

We now briefly describe the particular results for the Warner potential when using the reduction procedure (Section 5). According to (35), the parameter $\Lambda$ in (22), (23) and the parameters $G_k^r$ ($k = 0,1,2$) in (24) are of the form:

$$\Lambda = \Delta l^2 /(\Delta l)^2 \, ; \quad G_0^r = -G_1^r = G_2^r = 2G. \tag{36}$$

Due to (36) the reduced Warner infinitesimal potential and related reduced extra stress take the respected expressions:

$$\rho F_w^r / G = tr\underline{\underline{E}}^2 - 2[tr(\underline{nn} \cdot \underline{\underline{E}}^2) - (1-\alpha)tr^2(\underline{nn} \cdot \underline{\underline{E}})], \tag{37}$$

$$\underline{\underline{\sigma}}_w^r /(2G) = \underline{\underline{E}} - [\underline{nn} \cdot \underline{\underline{E}} + \underline{\underline{E}} \cdot \underline{nn} - 2(1-\alpha)\underline{nn}tr(\underline{\underline{E}} \cdot \underline{nn})]. \tag{38}$$

An important feature of equations (37) and (38) is that except parameter $\alpha$, they do not contain material parameters describing orientation. These parameters are only presented in kinematic equations (22) and (23) via the parameter $\Lambda$ expressed in this case in (36). Additionally, equations (37) and (38) show the existence of the soft deformation modes, since due to the second (chain) equality (35), the condition of marginal instability (30) is fulfilled.

## 8. Conclusions

1. The general case of incompressible non-classical elasticity with small deformations, and small three dimensional, internal and body rotations is analysed for uniform nematic elastomers. It was shown that the Frank's nematic contribution to this type of elasticity is negligible for space scales more than $10^3$ *nm*. Since initial and actual (in deformed state) values of director are related via internal rigid rotations, the state variables include small elastic deformations and relative rotations, with known initial distribution of director field. Additionally, a small change in the scalar order parameter was considered as an independent state variable. In this case, the free energy density is presented as a quadratic form with respect to elastic strains, internal rotations, and variation of scalar order parameter, with coefficients depending on the orientation of initial director. De Gennes (1980) originated such a formulation for LC elastomers, however, neglecting the effect of internal spin and scalar order parameter. It was shown that in the static situations, the effect of scalar order parameter just changes a material constant, determined for a less general situation where the change in scalar order parameter due to mechanical field is neglected.

2. A complete analysis of thermodynamic stability of the general potential resulted in finding necessary and sufficient conditions of stability, imposed on material parameters in the free energy density. Because of the presence of relative rotations, the stress tensor calculated using the free energy density is generally non-symmetric.

3. As follows from the kinetics of internal rotations, in the absence of external fields and neglecting the inertia effects of internal rotations and couple stresses, stress tensor is symmetric. Such a situation is common for equilibrium of anisotropic elastic solids. In this case, nullifying the possible anti-symmetric part stress tensor allows expressing relative rotations via the elastic strain and initial value of director. Substituting this relation into the free energy density and symmetric part of stress represents both of them via elastic strain and initial (or actual) value of director, with preserving the potential relation for extra stress and stability conditions. This procedure of *reduction* of free energy and stress greatly simplifies all the formulations and may be used for studies of relaxation phenomena.

4. Using the reduced formulation for the free energy and stress, a general condition for existence of soft deformations for weak nematic elasticity was found and analysed. This effect of marginal instability, commonly explained before only for the Warner potential, demonstrates a spontaneous transformation of elastic energy into rotational energy, without expenditure of total energy.

5. As an example, Warner potential that have been formulated for large elastic deformations of nematic elastomers, was analysed for small elastic deformations and rotations, and also for small change in the scalar order parameter. It was shown that the reduced potential and stress tensor do not contain any material parameters characterizing orientation, except for one that characterizes the effect of scalar order parameter. The expression for reduced potential clearly demonstrates the existence of soft deformations, because it always satisfies the condition of marginal instability.

# References

Abramchuk S. S., Khokhlov A. R., 1987. Molecular theory of high elasticity of polymer networks, with an accounting for orientational ordering of units. Dokl. Akad. Nauk USSR (in Russian) 297, 385-388.

Allen S.J., de Silva C.N., 1966. A theory of transversely isotropic fluids. J. Fluid Mech., 24, 801-821.

Anderson D. R., Carlson D. E., Fried E., 1999. A continuum-mechanical theory for nematic elastomers. J. Elasticity, 56, 33-58.

Bladon P., Terentjev E. M., Warner M., 1993. Transitions and instabilities in liquid-crystal elastomers. Phys. Rev. E., **47**, R3838-R3840 (1993).

Brand H. R., 1989. Electromechanical effects in cholesteric and chiral smectic liquid-crystalline elastomers. Macromol. Chem. Rapid Commun. 10, 441-445.

Brand H. R., Pleiner H., 1994. Electrohydrodynamics of nematic liquid crystalline elastomers. Physica A, 208, 359-372.

Cosserat, E. & F., 1909. Theorie des corps deformables, Paris: Hermann.

de Gennes P.G., 1980. Weak nematic gels. In: Liquid crystals of one-and two-dimensional order, Helfrich W. and Heppke G. Eds., Springer, Berlin, 231-237.

de Groot, S.R., Mazur, P., 1962. Non-equilibrium thermodynamics, North-Holland, Amsterdam, Ch.12.

Ericksen, J.L., 1960. Transversely isotropic fluids. Kolloid Z, 173, 117-122.

Frank, F.C., 1958. On the theory of liquid crystals. Discuss. Faraday Soc., 25, 19-28.

Golubovich L., Lubensky T.C., 1989. Nonlinear elasticity of amorphous solids. Phys. Rev. Lett., 63, 1082-1085.

Halperin A., 1986. Mean field theory of nematic order in gels. J. Chem. Phys., 85, 1081-1084.

Leslie, F. M., 1968. Some constitutive equations for liquid crystals, Arch. Rational Mech. Analysis, 28, 265-283.

Olmsted P.D., 1994. Rotational invariance and Goldstone modes in nematic elastomers and gels. J. Phys. II , 4, 2215-2230.

Oseen, C.W., 1925. Neue Grundlegung der Theory der anisotropen Flussigkeiten. Arkiv Matematik, Astronomi, Fysik. 19A, 16-19.

Terentjev, E.M., 1993. Phenomenological theory of non-uniform nematic elastomers: free energy of deformations and electric field effects. Europhys. Lett., 23, 27-32.

Terentjev E. M., Warner M., Verwey G. C., 1996. Non-uniform deformations in liquid crystalline elastomers. J. Phys. II , 6, 1049-1060.

Terentjev E.M., Warner M., 2001. Linear hydrodynamics and viscoelasticity of nematic elastomers, Eur. Phys. J. E., 4, 343-353.

Warner M., 1999. J. Mech. Phys. Sol., New elastic behaviour arising from the unusual constitutive relation of nematic solids. 47, 1355-1377.

Warner M., Bladon P., Terentjev E. M., 1994. "Soft elasticity" – deformation without resistance in liquid crystal elastomers". J. Phys. II, 4, 93-102.

Warner M., Gelling K. P., Vilgis T. A., 1988. Theory of nematic networks. J. Chem. Phys., 88, 4008-4013.

Warner M., Wang X. J., 1991. Elasticity and phase behavior of nematic elastomers, Macromolecules, 24, 4932-4941.

Weilepp J., Brand H. R., 1996. Director reorientation in nematic-liquid-single-crystal elastomers by external mechanical stress. Europhys. Lett. 34, 495-500.